\documentclass[aps,prl,twocolumn,showpacs,superscriptaddress]{revtex4}
\usepackage{amsfonts}
\usepackage{txfonts}
\usepackage{epsfig}

\def\be{\begin{equation}} \def\ee{\end{equation}}
\def\bea{\begin{eqnarray}} \def\eea{\end{eqnarray}}

\def\nn{\nonumber}

\begin{document}
\title{Topological order parameters for interacting topological insulators}

\author{Zhong Wang}

 \affiliation{Department of Modern Physics, University of
Science and Technology of China, Hefei, 230026, P. R. China}

 \affiliation{Department of Physics, McCullough Building, Stanford
University, Stanford CA 94305-4045}

\author{Xiao-Liang Qi}

\affiliation{Microsoft Research,
Station Q, Elings Hall, University of
California, Santa Barbara, CA 93106}

\affiliation{Department of Physics, McCullough Building, Stanford
University, Stanford CA 94305-4045}

 \author{Shou-Cheng Zhang}

  \affiliation{Department of Physics, McCullough Building, Stanford
University, Stanford CA 94305-4045}

\begin{abstract}
We propose a topological order parameter for interacting topological
insulators, expressed in terms of the full Green's functions of the
interacting system. We show that it is exactly quantized for a time
reversal invariant topological insulator, and it can be
experimentally measured through the topological magneto-electric
effect. This topological order parameter can be applied to both
interacting and disordered systems, and used for determining their phase diagrams.
\end{abstract}
 \maketitle

Recently, topological insulators(TI) have attracted great attention
in condensed matter physics\cite{qi2010,moore2010,hasan2010}.
Historically, the concept of the time reversal ($\cal T$) invariant
TI has been developed along two independent routes\cite{qi2010}.
The topological field theory (TFT) approach was first introduced by
Zhang and Hu\cite{Zhang2001},  who constructed a microscopic
model of the $\cal T$-invariant TI in four spatial
dimensions($4+1$d), and showed that the effective TFT is described
by the $4+1$d Chern-Simons(CS) action\cite{bernevig2002}. In
contrast to the $\cal T$ breaking CS action in $2+1d$, the CS action
in $4+1d$ preserves the $\cal T$ symmetry. It is now understood that
this state is the fundamental $\cal T$-invariant TI state, from
which both the three and the two dimensional $\cal T$-invariant TIs
can be derived\cite{qi2008,Kitaev2009,ryu2009}. Using the simple
procedure of dimensional reduction, the TFT for three and two
dimensional $\cal T$-invariant TIs has been
constructed\cite{qi2008}. Independently, the topological band theory
(TBT) was first developed starting from the pioneering work of Kane
and Mele\cite{kane2005b}, who first proposed a $Z_2$
topological invariant within the non-interacting band theory. This
non-interacting $Z_2$ topological invariant has been generalized to
the three dimensional TIs\cite{fu2007b,moore2007,Roy2009}. The TFT
is generally valid for systems with interactions and disorder;
recently, it has been shown that the TFT reduces exactly to the the
TBT in the non-interacting limit\cite{wang2009}. The TFT has been
further developed recently\cite{qi2009,essin2009, karch2009,
rosenberg2010}.

TIs are widely believed to be a new state of quantum matter. Since all states of matter in Nature, including band insulators, are necessarily interacting, it is important to formulate a general definition of TIs which is valid for general interactions and disorder.
This is especially important since an explicit counter-example in
one dimension has been constructed where a non-interacting
topological state becomes unstable against
interactions\cite{fidkowski2009}. After the theoretical prediction
and experimental discoveries of the weakly interacting
TIs\cite{qi2010,moore2010,hasan2010}, it would be most interesting
to investigate topological Mott insulators where the interaction
plays an essential
role\cite{raghu2008,shitade2009,pesin2009,zhang2009a,seradjeh2009,
li2010,dzero2010,rachel2010}. In the case where strong electronic
correlations play a crucial role, it is essential to define a
general topological order parameter which can determine the phase
diagram of these systems.

The TFT approach defines the generally interacting TI in terms of a topological term in the effective electromagnetic action. The $\theta$ angle, or equivalently the electromagnetic polarization $P_3$, can only take two discrete values in a system with ${\cal T}$ symmetry. In this work, we give an explicit formula of $P_3$ which can be evaluated in a generally interacting system. Our explicit formula for the topological order parameter $P_3$ satisfies the following general criteria: (1) It is well defined in the presence of interaction and disorder. (2) It takes quantized values, invariant against small changes of parameters in the model Hamiltonian or in experiment. (3) It is experimentally measurable. The TBT is based on the single particle band states, and
can not be easily generalized to interacting systems. On the other
hand, the TFT is valid for systems with general interactions and
disroder, and the quantized magneto-electric polarization can be
directly measured
experimentally\cite{qi2008,qi2009,tse2010,maciejko2010}. The central
result of this paper is the topological order parameter of the
$3+1$d TI defined as
\begin{figure}
\includegraphics[width=8.0cm, height=3.5cm]{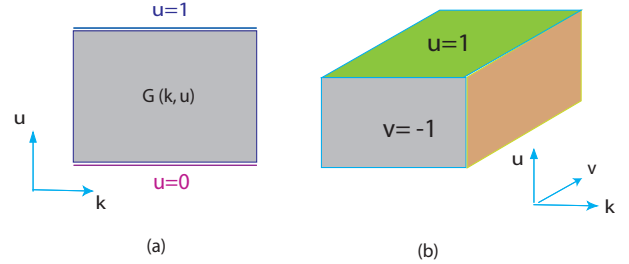}
\caption{Extension of $G(k)$ to higher dimensional manifolds $(k,u)$
and $(k,u,v)$. (a)Extension of $G(k)$ to $G(k,u),\, u \in [0,1]$ in the definition of $P_3$,
with $G(k,u=0)= G(k)$. The $3+1$d momenta space is illustrated as horizontal 1d
interval. (b)Extension of $G(k)$ to $G(k,u,v)$ with $G(k,u=v=0)=
G(k)$ in the definition of $P_2$.}
\label{wzw}
\end{figure}
\bea P_{3} &=& \frac{\pi}{6} \int_{0}^{1}du \int
\frac{d^{4}k}{(2\pi)^{4}} \textrm{Tr} \epsilon^{\mu \nu \rho \sigma}
[G\partial_{\mu}G^{-1} G\partial_{\nu}G^{-1} \nonumber
\\  &\,& \times G\partial_{\rho}G^{-1} G\partial_{\sigma}G^{-1}
G\partial_{u}G^{-1}] \label{P3} \eea in which $G=G(k,u);\,
k=(k_0,{\bf k})$. The momentum variables ${\bf k} = (k_1,k_2,k_3)$
are integrated over the Brillouin zone and frequency $k_0$ is
integrated over $(-\infty,+\infty)$. $G(k,u=0) \equiv G(k)$ is the
full imaginary time single-particle Green's function of the
interacting system, and $G(k,u)$ for $u\neq 0$ is a smooth extension
of $G(k,u=0)$ [see Fig.(\ref{wzw}a)], with a fixed reference value
$G(k, u=1)$ corresponding to Green's function of a topologically
trivial insulating state. It is convenient to choose $G(k, u=1)$ as
a diagonal matrix with $G_{\alpha \alpha} = (i k_{0} - \Delta)^{-1}
$ for empty bands $\alpha$ and $G_{\beta \beta} = (i k_{0} +
\Delta)^{-1} $ for filled bands $\beta$, where $\Delta
> 0$ is independent of ${\bf k}$. Even though $P_{3}$
is a physical quantity in three dimensions, we introduced a
Wess-Zumino-Witten (WZW)\cite{witten1983} type of extension
parameter $u$ in its definition. Essentially, we borrowed the ideas
of WZW term in coordinate space to the frequency and momentum space.
Similar to the WZW term, we shall show that $P_3$ is only
well-defined modulo an integer, and it can only take quantized value
of $0$ or $1/2$ modulo integer for an $\cal T$-invariant insulator.
The TFT describes the physical response of an insulator in terms of
the effective electromagnetic action\cite{qi2008} \bea S_{\rm
eff}&=& S_{\rm Maxwell}+S_{\rm topo}\nonumber \\ &=& \int d^3xdt
[\frac1{16\pi}F_{\mu\nu}F^{\mu\nu}+\frac{\theta\alpha}{32\pi^{2}}
\epsilon^{\mu\nu\sigma\tau} F_{\mu\nu}F_{\sigma\tau}]\label{s3d}
\eea where $\theta$ can only take values of $0$ or $\pi$ for a $\cal
T$-invariant insulator satisfying periodic boundary condition. We
show that the topological order parameter $P_3$ enters the physical
response function by the identity: \bea \theta =2\pi P_3
\label{response} \eea Therefore, our topological order parameter
$P_3$ satisfies all three main criteria discussed above.
Furthermore, the full Green's function entering $P_3$ can be
directly computed by quantum many-body techniques, such as exact
numerical diagonalization, quantum Monte Carlo method, dynamical
mean field theory etc.

{\it The fundamental $\cal T$-invariant TI in 4+1d:} Even though we
are mostly interested in the 3+1d physical space, the presence of
the WZW extension parameter $u$ hints that the topological structure
of $P_3$ is inherited from the $4+1$d TI \cite{Zhang2001}. The
concept of a $\cal T$ invariant TI is most naturally formulated in
$4+1$d. All the lower dimensional $\cal T$-invariant TIs can be
easily obtained from this fundamental state by a simple procedure of
dimensional reduction. For this reason, we discuss the fundamental
$\cal T$-invariant TI first in $4+1$d. The effective TFT for the
4+1d TI is given by a CS term \bea S_{\rm
eff}=\frac{C_2}{24\pi^2}\int
d^4xdt\epsilon^{\mu\nu\rho\sigma\tau}A_\mu\partial_\nu
A_\rho\partial_\sigma A_\tau  \label{4dCS} \eea Under the time
reversal transformation, ${\bf A} \rightarrow -{\bf A},$ $A_0
\rightarrow A_0$, therefore, we see that this term is explicitly
$\cal T$-invariant.

We first show that $C_2$ is quantized to be integer for insulating
system without ground state degeneracy on a 4d torus $T^4=T_{12}\times T_{34}$. Suppose we penetrate
a flux quanta $\phi = 2\pi$ into
$T_{34}$, with field strength $F_{34}= \partial_3 A_4 - \partial_4
A_3 = \phi/L^2$. On the 2d torus $T_{12}$, the 4+1d CS term is reduced to  $S_{{\rm eff}} = \frac{\phi}{2\pi}
\frac{C_2}{4\pi} \int dx_1dx_2dx_0 \epsilon^{\mu \nu \rho} A_{\mu}
\partial_{\nu} A_{\rho} = \frac{C_2}{4\pi}
\int dx_1dx_2dx_0 \epsilon^{\mu \nu \rho} A_{\mu}
\partial_{\nu} A_{\rho}$
with $\mu, \nu, \rho = 0,1,2$. This is just a 2+1d CS term. We consider an adiabatic evolution
of gauge potential $(A_1(t),A_2(t))$ along a rectangular path $C$:
$(0,0) \rightarrow (2\pi/L,0) \rightarrow (2\pi/L,2\pi/L)
\rightarrow (0,2\pi/L) \rightarrow (0,0)$, and we have Berry phase
given by CS term as $S_{{\rm eff}}=  2\pi C_2$. Because
$A_i$ is gauge equivalent to $A_i + 2\pi/L$, the system is actually
adiabatically evolving on a torus parameterized by
$(A_1,A_2),\,A_1,A_2 \in [0,2\pi/L]$. With the path $C$ enclosing
the entire torus surface, we have the Dirac quantization condition $
C_2= \textrm{integer}$. The only assumption in this argument is that
we have a unique ground state; otherwise the Berry phase is
generally non-abelian\cite{niu1985} and our argument fails. From the
integer quantization of $C_2$, we reach the conclusion that $C_2$ is
unchanged when the the Hamiltonian is tuned smoothly and the energy
gap remains open.

\begin{figure}
\includegraphics[width=5.5cm, height=3.5cm]{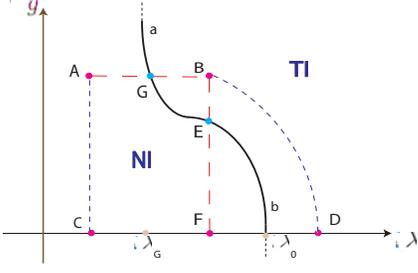}
\caption{Phase diagram on the $(\lambda, g)$ plane. The dark curve
$ab$ is the phase boundary seperating the normal insulators(NI) and
topologically nontrivial insulators(TI). The ground states are
gapped everywhere except on $ab$. The true parameter space is in
fact infinite dimensional, but this 2d diagram can illustrate the
main features. This diagram is applicable in both 4+1d and 3+1d. }
\label{connection}
\end{figure}

Now we turn to the topological order parameter for such $4+1$d TIs
with general interaction. We define \bea N_4 &\equiv&
\frac{\pi^2}{15} \int \frac{d^{5}k}{(2\pi)^{5}}
\textrm{Tr}[\epsilon^{\mu \nu \rho \sigma \tau}
G\partial_{\mu}G^{-1} G\partial_{\nu}G^{-1} G\partial_{\rho}G^{-1}
\nn \\ &\,& \times G\partial_{\sigma}G^{-1} G\partial_{\tau}G^{-1}]
\,   \label{N2} \eea where the partial derivative is taken with
respect to the momenta $k_{\mu}=(k_0,k_1,k_2,k_3,k_4)$, and $G=G(k)$
is the full Green's function. From the spectral representation of
$G(k)$, it can be shown that $G(k)$ is a smooth function of $k$ when
the energy gap is nonzero. Under a smooth change of physical
parameter, and the associated smooth change of $\delta G(k)$, the
variation $\delta N_4$ vanishes, as proven in Eq. (C1) of
Ref.~\cite{qi2008}. Therefore, $N_4$ is a topological invariant.
Next we show that $N_4$ is always an integer. Generally, $G(k)$
defines a map from the five dimensional $k$ space to the space of
non-singular Green's functions, belonging to the group ${\rm
GL}(n,{\rm C})$, whose homotopy group is labeled by an integer: \bea
\pi_5({\rm GL}(n,{\rm C}))= \mathbb{Z} \label{homotopy} \eea which
is exactly $N_4$. Here $n\geq 3$ is the number of bands. Finally, we
show that the identity \bea C_2= N_4 \label{csresponse} \eea which
is a $4+1$d analog of Eq.(\ref{response}), holds for general
interacting systems. To be specific, we consider a typical phase
diagram shown in Fig.(\ref{connection}) for an interacting
Hamiltonian $H=H_0(\lambda)+H_1(g)$, where $H_0$ is the
non-interacting part including terms such as $t_{ij}c_i^\dag c_j$,
and $H_1$ is the electron-electron(e-e) interaction part including
terms such as the Hubbard interaction $g n_{i\downarrow}
n_{i\uparrow}$. These two parts $H_0(\lambda)$ and $H_1(g)$ are
determined by single particle parameters
$\lambda=(\lambda_1,\lambda_2,\cdots)$ and coupling constants
$g=(g_1,g_2,\cdots)$ respectively. When $(\lambda, g)$ are smoothly
tuned, the ground state evolves smoothly so long as the energy gap
remains open, and therefore both $N_4$ and $C_2$ remain unchanged,
as has been discussed. Only when the gap closes, the full Green's
function $G$ becomes singular, and $N_4$ and $C_2$ can both change,
as indicated by the curve $ab$ in Fig.(\ref{connection}).
Arbitrarily picking a gapped state $A$ in the phase diagram, we can
find a path $AC$ connecting $A$ to a non-interacting state $C$,
without crossing the phase boundary $ab$. Now we have $N_4(A) =
N_4(C)$ and $C_2(A)=C_2(C)$. In addition, because $C$ is a
non-interacting state, $C_2$ can be simply calculated from a single
Feynman diagram, which gives the result $C_2(C)=N_4(C)$
\cite{qi2008,niemi1983,golterman1993,volovik2002}. Therefore, we
have $N_4(A)=C_2(A)$ for generally interacting ground state $A$.

Our discussion so far can be straightforwardly generalized to CS
insulators in $2n+1$d for integer $n$. In particular, we give here
the topological order parameter $N_2$ for the two dimensional
quantum anomalous Hall insulator\cite{haldane1988,qi2005}: \bea N_2
= \frac{\pi}{3} \int \frac{d^3k}{(2\pi)^3} {\rm Tr}[\epsilon^{\mu
\nu \rho} G\partial_\mu G^{-1}G\partial_\nu G^{-1} G\partial_\rho
G^{-1}] \eea This formula generalizes the TKNN
formula\cite{thouless1982} to the interacting systems, and is
equivalent to the Chern number formula with twisted boundary
condition\cite{niu1985}. It could be useful to determine the phase
diagram of an quantum anomalous Hall insulator induced by
interactions\cite{raghu2008}. This formula could also be used to define the chiral topological superconductor\cite{volovik2003}, where $G$ is regarded as the Green's function of the
BdG quasi-particles.

{\it The $\cal T$-invariant TI in 3+1d:} Now we turn attention to $\cal T$-invariant insulators in 3+1d. Following the logic of dimensional reduction\cite{qi2008}, we propose a topological order parameter defined by Eq.(\ref{P3}).

Now we show $P_3$ is quantized to be integer or half-integer.
First,
it can be obtained that $G(k, u=0)$ has the $\cal T$ property
$G(k_{0},-{\bf k}, u=0)=T G(k_{0},{\bf k}, u=0)^{T} T^{\dag}$, where
$^{T}$ means transposition and $T$ is the time-reversal matrix
satisfying $T^{\dag}T= 1$ and $T^{\ast}T= -1$. Given an extension
$G(k, u)$, with $u\in [0,1]$, we consider a mirror extension,
defined as $\tilde{G}(k,u)$, with the property $\tilde{G}(k_{0},{\bf
k}, u)=T G(k_{0},{\bf -k},- u )^{T} T^{\dag}, u \in [-1,0]$, for
which $\tilde{P_3}$ is defined similarly to Eq.(\ref{P3}), but with
$u$ integrated over $[-1,0]$ instead of $[0,1]$. It can be checked
that $P_3 = \tilde{P_3} $ for a mirror pair of extensions $G$ and
$\tilde{G}$, therefore, we have \bea 2P_3 &=& P_3+\tilde{P_3}
\nonumber \\
 &=& \frac{\pi}{6} \int_{-1}^{1}du \int
\frac{d^{4}k}{(2\pi)^{4}} \textrm{Tr} \epsilon^{\mu \nu \rho
\sigma} [\hat{G}\partial_{\mu}\hat{G}^{-1}  \nonumber \\
 &\,& \times \hat{G}\partial_{\nu}\hat{G}^{-1} \hat{G}\partial_{\rho}
 \hat{G}^{-1} \hat{G}\partial_{\sigma}\hat{G}^{-1} \hat{G}\partial_{u}
 \hat{G}^{-1}]
 \nonumber \\
&=&  n \label{2P3}
 \eea where $n$ is an integer, and
$\hat{G}(k,u)=G(k,u)$ when $u \in [0,1]$ while
$\hat{G}(k,u)=\tilde{G}(k,u)$ when $u \in [-1,0]$. The integral in
Eq.(\ref{2P3}) is an integer because
$\hat{G}(k,u=1)=\hat{G}(k,u=-1)$ are both the same reference Green's
function, so we can identify the manifolds $(k,u=-1)$ and $(k,u=1)$
as the same manifold, and we have an integral over a torus in ${\bf
k}$ space($k_0$ is still integrated over $(-\infty,\infty)$). From
Eq.(\ref{2P3}), we know that $P_3=n/2$ is quantized to be integer or
half-integer. Therefore, its variation under a infinitesimal change
of $G$ vanishes, just like $N_4$. It is well known that the WZW
terms have integer ambiguity\cite{witten1983}. If we choose two
different WZW extensions, their difference is generally given by the
homotopy class in Eq.(\ref{homotopy}), which is an integer. The
integer ambiguity of $P_3$ translates into the periodicity of
$\theta$ under a shift of $2\pi$, by the identification of Eq.
(\ref{response}), which can be obtained in similar way to Eq.(\ref{csresponse}). Therefore, $P_3$ is a $Z_2$ topological order
parameter for $3+1$d $\cal T$-invariant interacting insulators.

Now we briefly discuss the effect of disorder. It is convenient to
consider a large but finite size insulator with periodic boundary
condition. To define $P_3$, we can use a twisted boundary
condition\cite{niu1985,Avron1985} with twisted phase $\theta_i \in
[0,2\pi],\, i=1,2,3$. The Green's function, as a function of
$\theta_i$, is now a matrix whose rank is proportional to the system
size instead of the number of bands. The $P_3$ for disordered
insulator is defined by simply replacing $k_i$ in Eq.(\ref{P3}) by
$\theta_i$. It can be shown that this new definition of $P_3$
reduces to Eq.(\ref{P3}) in the absence of disorder. The analysis in
the interacting insulator applies to the disordered insulator as
well, and we have similar phase diagram as Fig.(\ref{connection}),
with $\lambda$ interpreted as the spin-orbit coupling and $g$
interpreted as the disorder strength.

Therefore, using the simple topological order parameter $P_3$
expressed in term of Green's function, we have a unified picture of
TI in the presence of interaction and disorder. The same discussion
applies to $4+1$d and $2+1$d,  where we have disorder-induced
quantum spin Hall states, which has been studied in
Ref.\cite{li2009,groth2009}.

{\it The $\cal T$-invariant TI in 2+1d:} The topological order
parameter in $2+1$d is similar to the $3+1$d case. The only
difference is that in $2+1$d we need two WZW extension parameters
$u$ and $v$ to define our topological order parameter (see
Fig.(\ref{wzw}b)). Given a $2+1$d insulating system with full
Green's function $G(k)$, we can extend $G(k)$ to a 2d torus
parameterized by $(u,v); \, u,v\in [-1,1]$, i.e. we define
$G(k,u,v)$ satisfying $G(k,0,0)=G(k)$, and $G(k_{0},{\bf k}, u, v)=
T G(k_{0},-{\bf k}, -u, -v)^T T^\dag$. The Green's functions at the
boundary $u=1$ and $v=1$ are fixed to be some reference value, which
can be chosen to be trivial, say with flat bands, in the same way as
in $3+1$d. The topological order parameter in 2+1d is defined as

\bea P_2 &=& \frac{1}{120}\epsilon^{\mu\nu\rho\sigma\tau
}\int_{-1}^{1}du\int_{-1}^{1}dv\int\frac{d^3k}{\left(2\pi\right)^3}{\rm
Tr}[G\partial_{\mu}G^{-1} \nonumber
\\ &\,& \times G\partial_{\nu}G^{-1}G\partial_{\rho}G^{-1}
G\partial_{\sigma}G^{-1}G\partial_{\tau}G^{-1} \nonumber
\\ &=& 0 \,\,{\rm or}\, \, 1/2 \,\,\,({\rm mod \, integer} ) \label{P2}
\eea where $\epsilon^{\mu\nu\rho\sigma\tau } $  is the
anti-symmetric tensor taking value 1 when the variables are ordered
as $(k_0,k_1,k_2,u,v)$. The cases $P_2=0$ and $P_2=1/2$ modulo
integer correspond to topologically trivial and nontrivial
insulators in 2+1d, respectively. This topological order parameter
$P_2$ is valid for interacting quantum spin Hall systems in 2+1d,
including states in the Mott regime\cite{raghu2008}.  $P_2$ can be
physically measured by the fractional charge at the edge of the
quantum spin Hall state\cite{qi2009a}. Analog of $P_3$ and $P_2$ in (1+1)D cannot be defined because its value (mod integer) would vary when homotopically nonequivalent WZW extensions are used.

In conclusion we have introduced topological order parameters for
$\cal T$-invariant TIs in four, three and two dimensions. These
topological order parameters are defined in terms of the full
Green's function, and apply to both interacting and disordered
systems. These order parameters take quantized values which are
stable against small changes of physical parameters, and they can be
measured directly in experiments. Throughout this paper, we assume
no ground state degeneracy. New fractionalized topological phases
can emerge when this fundamental assumption is
removed\cite{bernevig2006a,levin2009b,maciejko2010b}.

This work is supported by the NSF under grant numbers DMR-0904264.
We thank Suk Bum Chung, Joseph Maciejko, Srinivas Raghu, Ramamurti
Shankar, Shao-Long Wan and Yong-shi Wu for helpful discussions. Z.W.
acknowledges the support of NSF of China(Grant No.10675108), CSC and SLAC
National Lab. X.L.Q acknowledges the support of
Microsoft Research Station Q.

\bibliography{interaction}

\end{document}